# Emission of coherent THz magnons in an antiferromagnetic insulator triggered by ultrafast spin-phonon interactions

E. Rongione[1,2], O. Gueckstock[3], M. Mattern[4], O. Gomonay[5], H. Meer[5], C. Schmitt[5], R. Ramos[6], E. Saitoh[6], J. Sinova[5], H. Jaffrès[1], M. Mičica[2], J. Mangeney[2], S. T. B. Goennenwein[7], S. Geprägs[8], T. Kampfrath[3], M. Kläui[5,9,10], M. Bargheer[4,11], T. S. Seifert[3,*], S. Dhillon[2], R. Lebrun[1,*]

[1] *Unité Mixte de Physique, CNRS, Thales, Université Paris-Saclay, F-91767 Palaiseau, France*
[2] *Laboratoire de Physique de l'Ecole Normale Supérieure, ENS, Université PSL, CNRS, Sorbonne Université, Université Paris Cité, F-75005 Paris, France*
[3] *Institute of Physics, Freie Universität Berlin, D-14195 Berlin, Germany*
[4] *Institut für Physik und Astronomie, Universität Potsdam, D-14476 Potsdam, Germany*
[5] *Institute of Physics, Johannes Gutenberg-University Mainz, D-55128 Mainz, Germany*
[6] *Institute for Materials Research and Center for Spintronics Research Network, Tohoku University, J-980-8577 Sendai, Japan*
[7] *Department of Physics, University of Konstanz, D-78457 Konstanz, Germany*
[8] *Walther-Meißner-Institut, Bayerische Akademie der Wissenschaften, D-85748 Garching, Germany*
[9] *Graduate School of Excellence Materials Science in Mainz (MAINZ), Staudingerweg 9, D-55128 Mainz, Germany*
[10] *Center for Quantum Spintronics, Department of Physics, Norwegian University of Science and Technology, N-7034 Trondheim, Norway*
[11] *Helmholtz-Zentrum Berlin für Materialien und Energie, Wilhelm-Conrad-Röntgen Campus, BESSY II, Albert-Einstein-Strasse 15, D-12489 Berlin, Germany*

*Corresponding authors: tom.seifert@fu-berlin.de, romain.lebrun@cnrs-thales.fr

## Abstract

Antiferromagnetic materials have been proposed as new types of narrowband THz spintronic devices owing to their ultrafast spin dynamics. Manipulating coherently their spin dynamics, however, remains a key challenge that is envisioned to be accomplished by spin-orbit torques or direct optical excitations. Here, we demonstrate the combined generation of broadband THz (incoherent) magnons and narrowband (coherent) magnons at 1 THz in low damping thin films of NiO/Pt. We evidence, experimentally and through modelling, two excitation processes of magnetization dynamics in NiO, an off-resonant instantaneous optical spin torque and a strain-wave-induced THz torque induced by ultrafast Pt excitation. Both phenomena lead to the emission of a THz signal through the inverse spin Hall effect in the adjacent heavy metal layer. We unravel the characteristic timescales of the two excitation processes found to be < 50 fs and > 300 fs, respectively, and thus open new routes towards the development of fast opto-spintronic devices based on antiferromagnetic materials.

## Introduction

Antiferromagnetic spintronics has recently become an important research field from both a fundamental viewpoint and its strong applicative potential [1,2]. Antiferromagnets (AFM) have key advantages linked to their magnetic ordering: they are insensitive to perturbative external magnetic fields, stray fields are absent, and magnon modal frequencies reach the terahertz (THz) regime [1–3]. This renders AFMs prime candidates for ultrafast spintronic devices [4] compared to their ferromagnetic counterpart. A recent work demonstrated the writing of AFM memory states with picosecond excitations [5]. In parallel, narrow band sub-THz detection has been achieved using spin-pumping in AFMs [6,7]. Another spintronic application is spintronic-based broadband THz emission that currently relies on ferromagnet/heavy metal heterostructures [8–13] and harnesses spin-to-charge-current conversion through the inverse spin Hall effect (ISHE). In this regard, AFM materials with their characteristic THz resonant modes could operate have been predicted to enable the development of narrowband THz spintronic emitters - THz nano-oscillators - controllable by spin-orbit torques [14]. A recent work observed broadband THz emission in AFM thin films, triggered by an off-resonant optical torque [15]. However, this work did not demonstrate the targeted narrowband emission of coherent THz magnons, which is highly desirable to fully functionalize spintronic THz emission.

Electrical switching of antiferromagnets recently highlighted that spin-orbit torques often compete with dominating thermo-magneto-elastic processes [16] and that thermally driven strain gradients can be used to reorient the antiferromagnetic domains [16]. Dynamic strain could, thus, be used to control antiferromagnetic states and potentially the AFM dynamics on ultrafast timescales [17–19]. Using this approach, one can envision to generate THz magnons using ultrafast strain gradients to achieve a time-dependent modulation of the exchange interaction.

In this paper, we report the combined generation of narrowband coherent THz emission centered at 1 THz and incoherent broadband THz magnons in NiO/Pt bilayers. For this purpose, we use femtosecond near-infrared laser pulses to either trigger direct light-spin interactions in (111) oriented films, that is, the inverse Cotton-Mouton effect (ICME) [20,21], or phonon-spin interactions in (001) oriented films through an ultrafast strain pulse that is generated upon heating of the metal layer [22–24]. We demonstrate that their respective efficiencies in generating THz spin-currents depend on the orientation and thickness of the AFM films. We identify the processes via the anisotropic or isotropic dependence on the pump polarization and show that the THz emission process arises in both cases from ultrafast spin-to-charge-current conversion in the heavy metal layer [15,25]. Finally, we quantify that the spin-current rise and decay times to be less than 50 fs, limited by the experimental resolution, for the direct light-spin coupling pathway, and to be more than 300 fs for the indirect spin-phonon excitation.

## Results and discussions

### Generation of broad- and narrowband THz emission in NiO(001)/Pt thin films

We first present the THz-emission signal in a NiO(001)(10nm)/Pt(2nm) bilayer (see Methods and Refs. [26,27]) when excited by 100 fs near-infrared (NIR) pulses under normal incidence (see **Fig. 1a**). Using THz emission time-domain spectroscopy (see Methods and Ref. [9]), we detect an emitted THz signal with two key features as shown in **Fig. 1b**: *i*) a very short THz pulse followed by *ii*) decaying oscillations with a periodicity of around 1 ps. In the frequency domain (inset of **Fig. 1b**), these two responses respectively correspond to *i*) a broadband contribution up to 3 THz (limited by the cutoff frequency of our detection crystal) and *ii*) a narrowband contribution centered at 1.1 THz.

The striking presence of oscillations at 1.1 THz is in direct agreement with the expected high-frequency magnon branch of NiO [28,29], which could only be observed previously in single crystals but not in thin films [15] due to the weak dipolar field generated by AFM magnon modes [28,30,31]. In application-relevant NiO/Pt bilayers, the spin current carried by propagating THz magnons can be converted into a transient THz charge-current through inverse spin-Hall effect in the Pt layer [15]. The presence of clear long-lived oscillations (of around 10 ps, compared to 30 ps in single crystals [28]), leading to the narrowband emission, indicates the low magnetic damping of the NiO thin films capped with Pt ($\alpha$ = 0.006). In optimized thin films (see **SM1**), the combined presence of broadband and narrowband contributions indicates that different emission mechanisms may contribute to the THz signal.

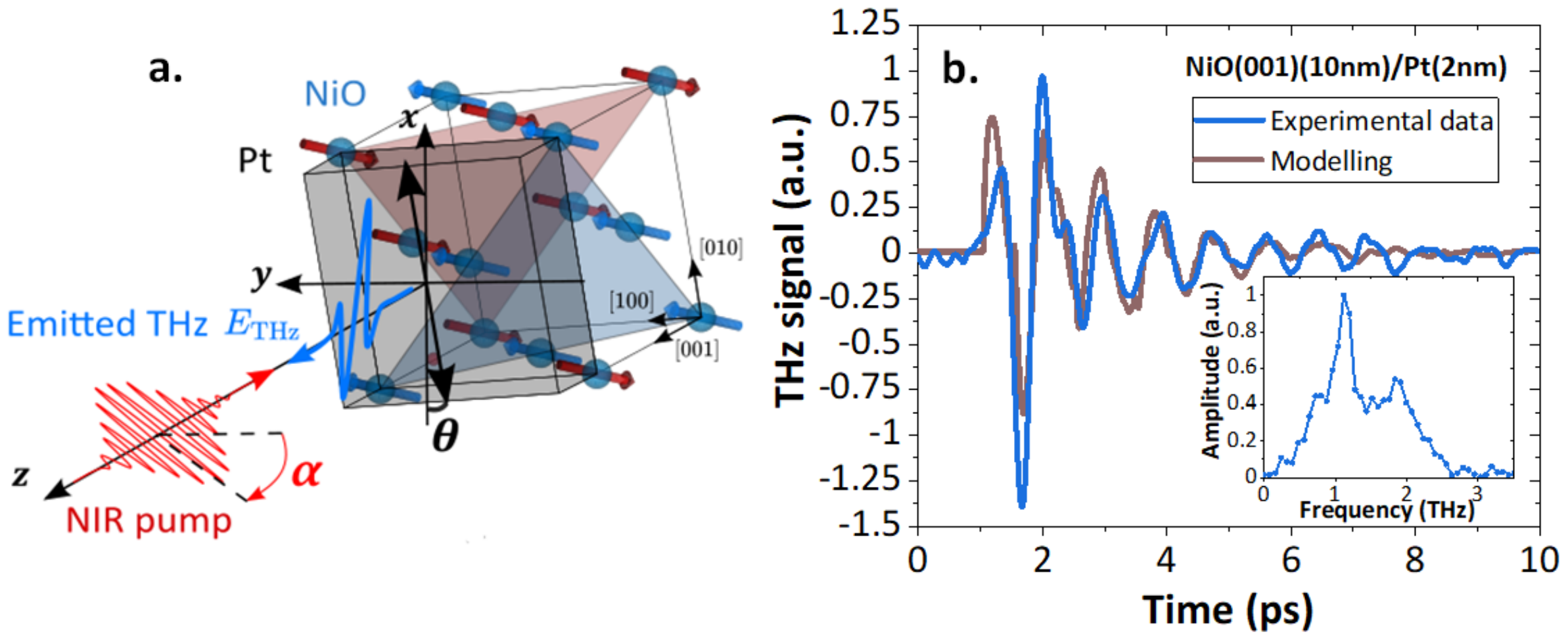

**Fig. 1. Laser induced coherent and incoherent THz emission from NiO/Pt bilayer.** (a) Schematic of the setup. Femtosecond NIR laser pulses excite NiO/Pt bilayers and the THz emission is collected at normal incidence in reflection geometry and detected using electro-optic sampling. $\theta$ corresponds to the in-plane sample orientation (defined from the sample edge [010] *i.e.* $\theta$ = 0° corresponds to the [010] sample edge along $\boldsymbol{x}$). $\alpha$ corresponds to the pump polarization angle (defined with respect to the $\boldsymbol{y}$ axis where $\alpha$ = 0°). In the lab frame ($\boldsymbol{x}, \boldsymbol{y}, \boldsymbol{z}$), $\boldsymbol{z}$ is the axis normal to the sample interface. For clarity, only the NiO magnetic sublattices without the oxygen atoms are represented. (b) Time domain THz emission from a low damping NiO(001)(10nm)/Pt(2nm) bilayer with the presence of oscillations about 1 ps of period (blue) and modelled THz response using magnetization dynamics simulation (brown). Inset: Fourier transform of the time domain signals.

We analyze the symmetries of the THz signal in **Fig. 2a** by measuring the dependence of the THz-emission signal in (001) oriented NiO films when rotating the in-plane angle $\theta$ ($\theta$ being the angle between the NiO [010] sample edge and the $\boldsymbol{x}$ axis, see **Fig. 1a**). The THz signal exhibits a uniaxial dependence, and the two THz emission lobes are opposite in phase as shown in **Fig. 2b.** Moreover, the emission axis at $\theta = \theta_0 \simeq 35°$ coincides with the orientation of the main *T*-domain in these thin films [32]. This observation indicates a THz emission originating from regions with a constant Néel vector orientation, that is, AFM monodomains of NiO with an area larger than the optical pump area (spot diameter is about 200 μm²) as confirmed by magneto-optic imaging [26,32] (inset of **Fig. 2b** and **SM2** [33]).

**Fig. 2c** shows that the THz-emission signal is independent of the linear pump polarization angle $\boldsymbol{\alpha}$ defined with respect to $\boldsymbol{y}$ for the NiO(001) film. This result suggests that light absorption in the Pt is the driving mechanism of ultrafast spin-current generation, in contrast to a recent report in (111) NiO thin films [15]. One must notice that there is no light absorption in the NiO as the pump photon energy of 1.5 eV is below the NiO bandgap of 4 eV, as confirmed by the linear fluence dependence that excludes multi-photon processes (see **SM3**). Moreover, the THz signal does reverse in phase upon reversal of the sample or by changing the capping layer from Pt to W (see **SM4**), as expected for a spin-to-charge (SCC) mechanism and not for a potential magnetic dipolar emission from the dynamics of the AFM moments. The THz-emission signal is also constant for external magnetic field of up to 200 mT (see **SM5**) in line with the large spin-flop fields in NiO (>10 T) [27].

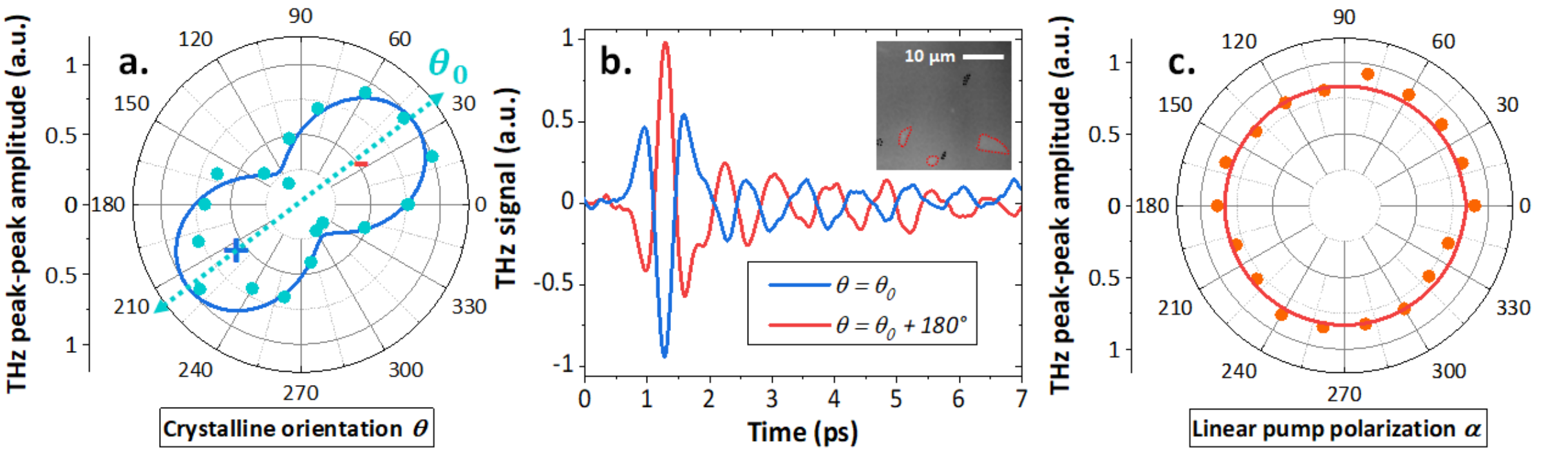

**Fig. 2. Symmetries of the THz-emission signal in NiO(001)(10nm)/Pt(2nm) bilayer.** (a) Dependence of the THz signal on the in-plane angle $\theta$, describing a rotation of the sample around the surface normal. THz emission shows a uniaxial behavior consistent with the presence of large $T$-domains (hundreds of microns in size) contribution. (b) THz emission for $\theta = \theta_0 \simeq 35°$ and $\theta = \theta_0 + 180°$ shows a sign reversal ($\pm$ label in panel a). The inset shows the magneto-optical birefringence imaging of an as grown NiO/Pt bilayer, presenting a majority of one $T$-domain orientation (grey contrast) with a small contribution of orthogonal minority $T$-domain (white contrast circled in red). The black spots are defects on the sample surface. (c) Dependence of the THz signal on the linear pump polarization (angle $\alpha$) showing an isotropic behavior in line with a thermal generation of magnons from light absorption in the Pt.

## Ultrafast mechanisms leading to THz magnonic spin-currents

To identify the THz emission mechanisms in NiO thin films, we measure the THz signal as a function of the NiO thickness (ranging from 5 nm to 110 nm) for two different growth orientations: either (001), as reported above, or (111), in which optical torques was recently reported [15], as shown in **Fig. 3a**. For (001)-oriented films, the THz emission is maximal for thinnest NiO and decreases to zero as the NiO thickness increases. The 1-THz oscillations in the time-domain can be resolved only for the optimized 5 and 10 nm thick (001) samples (see **SM1**). The unexpected thickness dependence of the THz signal can be associated with a relaxation of the static strain in thicker NiO films [26], together with potential destructive interferences of the generated magnonic current (see **SM6**). On the contrary, NiO(111) films exhibit a sizeable THz emission even for a large thickness of 110 nm unlike the (001) series. These observations point towards different excitation mechanisms depending on the orientation of the NiO films [15,20].

In **Fig. 3c-d,** we present the effect of the pump polarization (linear *vs.* circular) on the THz generation. Notably, we confirm the presence of two different behaviors for (001) and (111) orientations. For NiO(111) films, the THz amplitude is reduced to half the signal with circular pump polarization as shown in **Fig. 3c.** This hints at a direct off-resonant optical torque excitation of the NiO modes, identified through the ICME [15,20,21], which is line with the dependence on the linear pump polarization (see **SM7**). This is in stark contrast with the behavior observed for (001) films, where we find a THz-emission signal that is independent of the pump helicity (**Fig. 2c** and **Fig. 3d**), which we assign to light absorption in the Pt film. The polarization analysis agrees with a purely thermal origin of the emission of coherent THz magnons for the (001)-oriented films. We thus find strong evidences of two distinct excitation mechanisms, mediated either by light absorption in the Pt layer or direct off-resonant optical processes (by ICME) in the NiO layer, that can both contribute to the THz excitation of AFM insulating thin films.

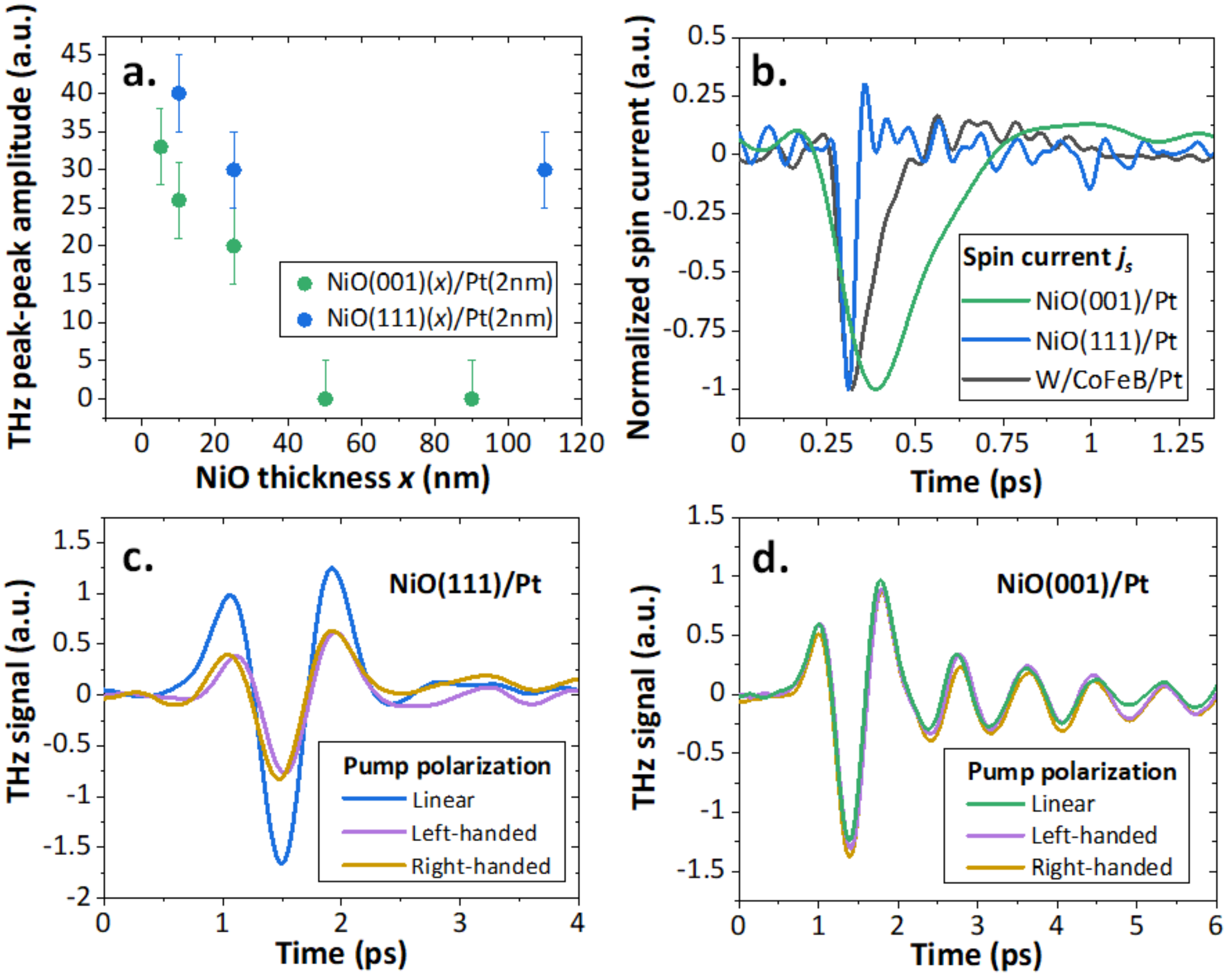

**Fig. 3. THz emission and dynamics for (001) *vs.* (111) NiO thin films.** (a) THz peak-to-peak amplitude for NiO(001) (green) and NiO(111) (blue) samples as a function of the NiO layer thickness $x$. (b) Extracted spin-current $j_s(t)$ entering the Pt layer generated in NiO (001) (10nm)/Pt(2nm) and NiO (111) (110nm)/Pt(10nm) samples compared to a fully metallic W(2nm)/CoFeB(1.8nm)/Pt(2nm) THz spintronic emitter. The ultrafast (<50 fs) rise and decay time of the spin-current for (111) sample is in line with inverse Cotton-Mouton effect mediated excitation while longer dynamics (>300 fs) in the (001) sample is in line with a thermal excitation mediated by optical absorption and phonons. 1 THz oscillations are not visible in this figure due to the time window scale. (c-d) Effect

of the pump polarization (linear or circular) on the THz-emission signal for (c) NiO(111) (25 nm)/Pt(2nm) and (d) NiO(001) (25 nm)/Pt (2 nm). Arbitrary zero-times are shifted for clarity.

Next, we explore the dynamics of these different mechanisms by extracting the temporal profile of the ultrafast spin-current using 20 fs pump pulses as illustrated in **Fig. 3b** (see Methods section and Ref. [34] for details). The three characterized samples are: a metallic CoFeB THz spintronic emitter [8] (TeraSpinTec GmbH) and two NiO/Pt bilayers with $(001)$ and $(111)$ orientations. The spin-current profile of the metallic THz emitter shows a typical rise time about 70 fs (associated with ultrafast demagnetization) [35] and decays within 250 fs (associated with the electron-spin relaxation time in metals) [35]. The NiO$(111)$/Pt sample presents an even faster spin-current rise time and decay time (< 50 fs). In contrast, the spin-current in NiO$(001)$(10nm) builds up over a much longer timescale of around 200-250 fs and relaxes over 300 fs. The THz spin-current dynamics further highlight the two generation mechanisms at stake in the NiO thin films. In case of $i$) $(111)$ NiO orientation, off-resonant Raman-type torques (as happening via ICME) acting on the magnetization are known to exhibit a quasi-instantaneous response, which explains the ultrafast rise time of the THz spin-current [20,21]. On the other hand, for $ii$) $(001)$ NiO orientation, the slower generation mechanism can be explained by an indirect thermal process that requires first the absorption of photons in Pt and subsequent processes that we will discuss next [34].

## Strain-wave driven THz magnonic excitation

Finally, we investigate the origin of the indirect excitation process in $(001)$ oriented NiO films by determining its spatio-temporal expansion (see **Fig. 4a**). We first measure the transient mean strain of the NiO layer $\varepsilon_{zz}^{\mathrm{NiO}}(t)$ using ultrafast reciprocal space mapping (URSM, see details in Methods and Refs. [36,37]). **Fig. 4b** presents the out-of-plane strain $\varepsilon_{zz}^{\mathrm{NiO}}(t)$ (grey dots) derived from time-resolved reciprocal space maps (see **SM8**) around the $(004)$ reflection peak of a NiO$(001)$(25nm)/Pt(2nm) thin film. One must notice that the very weak (004) peak signal prevents us from measuring thinner NiO films. The temporal strain response is characteristic of a bipolar strain wave driven by the ultrafast expansion of Pt upon optical heating, with a leading compression and a trailing expansion [23]. The strain wave is launched by the stress from electrons and phonons in Pt induced by the pump absorption by the electrons and the subsequent distribution of energy to the phonons (coupling time about 500 fs). From the initial state pictured in **Fig. 4c**, we observe three characteristic steps in the time-domain response: $i$) a compression of the NiO lattice up to 1.5-2 ps after pump pulse arrival due to the ultrafast expansion of Pt lattice upon optical heating (**Fig. 4d**). Then, we observe $ii$) a maximum expansion of NiO at 4 ps when the compressive part of the strain wave leaves NiO towards the substrate (**Fig. 4e**) and finally $iii$) a quasi-static (very slowly relaxing) expansion of NiO after 6 ps.

We then model the transient strain of NiO (blue line in **Fig. 4b**) using a simple one temperature model (and the Python toolbox *udkm1Dsim* [38]). We can thus get further insights into the dynamic of the system by calculating the spatio-temporal strain $\varepsilon_{zz}(z,t)$ and the temperature increase (see details in **SM8**). The total modeled strain in **Fig. 4b** (blue line) matches well the experimental results. It consists of an intrinsic heat expansion of NiO due to incoherent phonons excited by the temperature increase after heat transport from Pt into NiO (green dashed line), which is superimposed by propagating strain waves (coherent phonons) driven by Pt. It highlights two phononic contributions to the lattice response within the first picosecond: heating of NiO at the interface and compression of the first nanometers of NiO by the strain wave. From our model, we estimate the amplitude of the compression wave front about $6 \times 10^{-6}$ in the THz emission experiment for a fluence of about 10 µJ.cm$^{-2}$ (1000 times smaller than in the URSM experiment). This amplitude is sufficient to launch precession of the out-of-plane magnon mode via magneto-striction [16,39,40] as discussed later.

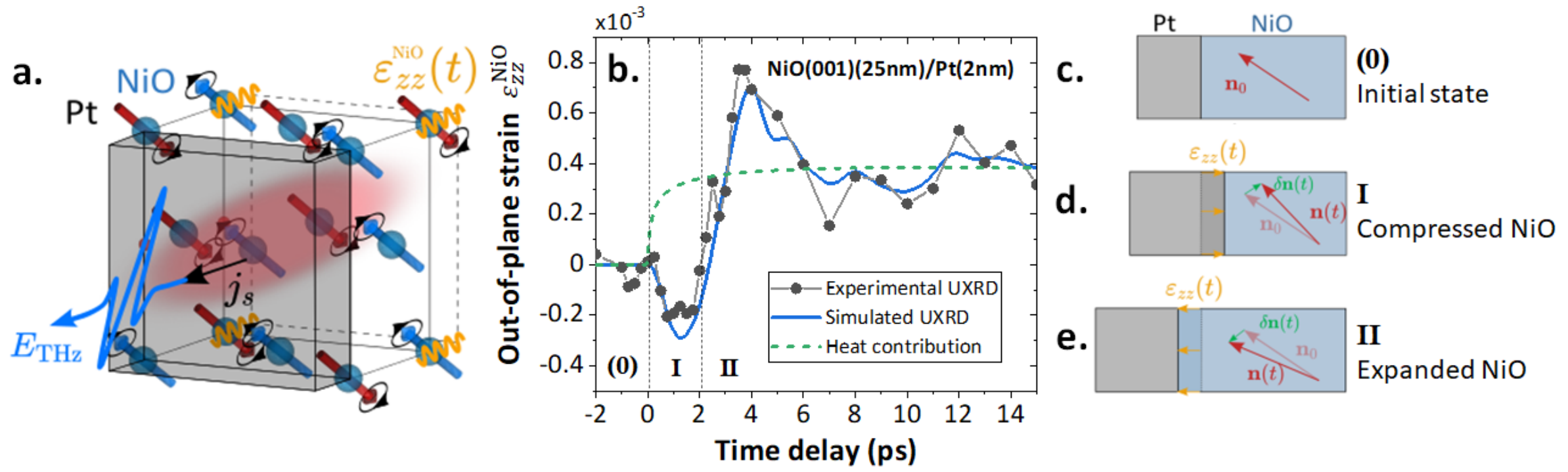

**Fig. 4. Ultrafast strain dynamics in NiO(001)/Pt leading to the THz emission.** (a) Sketch of the ultrafast spin-phonon interactions. The optical pump pulse is absorbed in Pt leading to an ultrafast lattice expansion of Pt following ultrafast electron-phonon coupling. A bipolar strain wave is thus launched into the NiO(001) layer. Via magneto-striction, this strain wave induces a deflection of the Néel order followed by out-of-plane oscillations, *i.e.* magnon excitations. A magnonic current flows into the Pt and leads to THz radiation by SCC**.** The temperature increase in the NiO additionally drives the spins out of equilibrium through the temperature dependence of the anisotropy field. (b) Out-of-plane mean strain $\varepsilon_{zz}^{\mathrm{NiO}}$ (grey) as a function of delay time mapped by ultrafast reciprocal space mapping (URSM) in a NiO(001)(25nm)/Pt(2nm) sample. The simulated strain (blue) includes contributions from the strain wave and from the quasi-static lattice expansion due to heating of NiO (dashed green line).

## Discussion

We finally discuss how different types of external torques $\boldsymbol{\Gamma}$ can induce the observed THz magnetization dynamics. Due to the wide spectrum of the pump laser pulse, the triggered magnetic dynamics can include a resonant excitation of THz magnon modes superimposed onto transient (non-resonant) oscillations in a wide frequency range (see **Fig. 1b**). We describe the magnetic state as a single-domain state of NiO with the Néel vector $\boldsymbol{n} \equiv \mathbf{M_1} - \mathbf{M_2}$ and solve the standard equations of motion [39]:

$$\boldsymbol{n_0} \times \left(\ddot{\boldsymbol{\delta n}} + 2\gamma_{\mathrm{AF}}\dot{\boldsymbol{\delta n}} - c^2\Delta\boldsymbol{\delta n} + \omega_{\mathrm{AF}}^2(T)\,\boldsymbol{\delta n}\right) = \gamma^2 H_{\mathrm{ex}}(\boldsymbol{n_0} \times \boldsymbol{\Gamma}), \quad (1)$$

where $\gamma$ is the gyromagnetic ratio, $H_{\mathrm{ex}}$ is the exchange field that keeps the magnetic sublattice moments antiparallel, $c$ is the limiting magnon velocity, $\omega_{\mathrm{AF}}$ is the circular frequency of the magnetic oscillations and $\gamma_{\mathrm{AF}}$ is the damping constant. Magnons are described as small deviations $\boldsymbol{\delta n}$ of the Néel vector from the equilibrium $\boldsymbol{n_0}$. Magnon spectra of NiO include two linearly polarized magnon branches [40], a low-frequency branch (with frequency ~180 GHz) and a high-frequency branch (with frequency ~1 THz). In the high-frequency branch, $\boldsymbol{\delta n}$ oscillates in the out-of-easy magnetic plane (either || [111] for NiO(111) and || [5 5 19] for NiO(001) due to a lattice distortion along [001] [26]), leading to a time-dependent dynamic magnetization $\boldsymbol{m} = \boldsymbol{n_0} \times \dot{\boldsymbol{\delta n}}/(\gamma H_{\mathrm{ex}})$ in the easy-plane. For the low-frequency branch, $\boldsymbol{\delta n}$ oscillates within the AFM plane with a dynamic magnetization $\boldsymbol{m}$ perpendicular to the AFM plane. The mode dynamics can be excited by an external torque $\boldsymbol{\Gamma}$ along $\boldsymbol{\delta n}$ multiplied by the exchange field $H_{\mathrm{ex}}$ (see **Eq. (1)**). A generated THz-emission signal $\boldsymbol{E}_{\mathrm{THz}}$ then emerges through the ISHE when the THz spin-current polarized along $\boldsymbol{m}$ has a finite projection onto the Pt plane. As such, only the high-frequency magnon branch can contribute to the THz generation for $(111)$ films while both modes could contribute for $(001)$ films.

For $(111)$ films, the measured THz-emission signals suggest that the main origin is an optical torque due to the direct interaction of the magnetic moments with the linearly polarized light (see **Fig. 3.c**), as also recently observed [15]. We can exclude the inverse Faraday effect, as it requires circular (or elliptic) polarization of light and consider the ICME [21] (see **SM6**). As the ICME originates from the off-resonant optical excitation of magnetic dynamics, the response time (spin-current rise time) of the NiO is defined by the linewidth of corresponding quantum processes of around the experimental resolution of 20 fs (as seen in **Fig. 3b**).

For (001) films, we find a largely pump-polarization independent THz-emission and thus consider how the torques induced by optical absorption in the Pt layer can generate a non-zero spin current from the NiO into the Pt layer. First, we can exclude the linear (in magnetization [41]) spin-Seebeck effect (SSE) [42,43], since it scales with the amplitude of the applied field for a compensated antiferromagnet [42](see also **SM9**). We therefore consider time-dependent spin-phonon interactions with two contributions. First, a coherent strain wave as measured in **Fig. 4** can trigger magnetization dynamics through the magneto-elastic effects [16,44,45], as discussed by Maldonado *et. al.* [46], which is a plausible mechanism for longitudinal out-of-plane strain along the $[001]$ direction. Injection of an ultrafast strain wave into NiO as observed in **Fig. 4** results in a bipolar modulation of the exchange interaction. Second, the lattice-heating due to incoherent phonons additionally contributes to the dynamical strain and lead to a change of magnetic anisotropy. These combined effects can be modelled by a torque $\Gamma \propto \lambda_{11} n_{0z} \varepsilon_{zz}(z,t)$, where $\lambda_{11}$ is the magneto-elastic constant that triggers the dynamic magnetization $\boldsymbol{m}(t)$ (with a deflection angle of about 0.3°, see **SM6**). This torque leads to a tilt of the easy magnetic plane, followed by a 1 THz out-of-plane oscillations of the Néel vector (see **SM6** for the details). On the contrary, for $(111)$ films, the ultrafast strain $\varepsilon_{zz}(t)$ induces oscillations of the Néel vector staying within the $(111)$ plane and thus no spin-current propagates towards the Pt layer. This key feature explains the difference between the two orientations of the NiO films [21]. Lastly, it should be noted that the symmetry of the $(001)$ films allows for a non-linear SSE [34] proportional to the temperature gradient at the NiO/Pt interface. This contribution has the same temporal signature as the lattice heating contribution in the Pt, with a spike signal just after the pump pulse followed by a decay in time (see **SM6**). Coherent strain, in contrast, induces oscillations of the dynamic magnetization $\boldsymbol{m}$ leading to oscillations of the injected spin current $j_s$. These combined effects contribute to the NiO magnetization dynamics and spin-current injection into Pt. The fit in **Fig. 1b** shows that the calculated combination of these effects (strain wave and temperature increase, see **SM6** for details) reproduces indeed qualitatively very well the experimental observations highlighting the different roads to generate THz magnonic current in an AFM insulator through spin-phonon interactions.

## Conclusion

We found strong evidence for two ultrafast mechanisms of magnonic excitations in the antiferromagnet NiO interfaced with a metallic Pt layer with largely different timescales. The THz emission appears via ultrafast off-resonant ICME in (111)-oriented NiO samples and/or spin-phonon interactions exciting the high-frequency magnon branch of NiO for (001) orientation. We also demonstrate by angular crystallographic THz emission mapping that the magnon generation mechanism is directly linked to the Néel vector orientation of NiO. This work opens routes towards AFM magneto-phononic and tunable THz narrowband emission.

## Acknowledgements

E. R., O. G., H. J., T. K., M. K., T. S. S., S. D. and R. L. acknowledge financial support from the Horizon 2020 Framework Programme of the European Commission under FET-Open grant agreement No. 863155 (s-Nebula). R. L. acknowledges financial support from the Horizon 2020 Framework Programme of the European Commission under FET-Open grant agreement No. 964931 (TSAR). H. J. and R. L. acknowledge financial support from the ANR project TRAPIST. M. M. and M. B. acknowledge financial support from the Deutsche Forschungsgemeinschaft (DFG, German Research Foundation) via No. BA 2281/11-1 Project-No. 328545488 – TRR 227, project A10. O G., H. M., C. S., J. S. and M. K. acknowledge financial support by the German Research Foundation DFG (CRC TRR 173 SPIN+X, projects 01, A03, A11, B02, #268565370) and TopDyn. M. K. acknowledges financial support by the DAAD (Spintronics Network #57334897 and 57524834) and the Research Council of Norway (Center of Excellence 262633 "QwSpin"). T. K. and T. S. S. acknowledge financial support from the German Research Foundation (DFG) through the collaborative research center SFB TRR 227 "Ultrafast spin dynamics" (Project ID 328545488, projects A05 and B02).